\documentclass[prd,aps,preprint]{revtex4-1}

\usepackage{epsfig}
\usepackage{graphicx,color}

\newcommand{\beq}{\begin{equation}}
\newcommand{\eeq}{\end{equation}}
\newcommand{\be}{\begin{eqnarray}}
\newcommand{\ee}{\end{eqnarray}}

\begin{document}

\title{
Singlet Fermionic Dark Matter with Dark $Z$
}

\author{ Dong-Won Jung$^1$ }
\email{dongwonj@korea.ac.kr}

\author{Soo-hyeon Nam$^1$}
\email{glvnsh@gmail.com}

\author{Chaehyun Yu$^1$}
\email{chyu@korea.ac.kr}

\affiliation{$^1$
Department of Physics,
Korea University, Seoul 02841, Korea
}

\author{Yeong Gyun Kim$^2$}
\email{ygkim@gnue.ac.kr}

\affiliation{$^2$
Department of Science Education,
Gwangju National University of Education, Gwangju 61204, Korea
}

\author{ Kang Young Lee$^3$ }
\email{kylee.phys@gnu.ac.kr}

\affiliation{$^3$
Department of Physics Education \& RINS
Gyeongsang National University, Jinju 52828, Korea
}

\date{\today}

\begin{abstract}

We present a fermionic dark matter model mediated by 
the hidden gauge boson.
We assume the QED-like hidden sector
which consists of a Dirac fermion and U(1)$_X$ gauge symmetry,
and introduce an additional scalar electroweak doublet field
with the U(1)$_X$ charge as a mediator.
The hidden U(1)$_X$ symmetry is spontaneously broken
by the electroweak symmetry breaking and
there exists a massive extra neutral gauge boson in this model
which is the mediator between the hidden and visible sectors.
Due to the U(1)$_X$ charge,
the additional scalar doublet does not couple to the Standard Model fermions,
which leads to the Higgs sector of type I two Higgs doublet model.
The new gauge boson couples to the Standard Model fermions 
with couplings proportional to those of the ordinary $Z$ boson 
but very suppressed, thus we call it the dark $Z$ boson.
We study the phenomenology of the dark $Z$ boson
and the Higgs sector, and 
show the hidden fermion can be the dark matter candidate.

\end{abstract}

\pacs{ }

\maketitle

\section{Introduction}

The Standard Model (SM) provides a consistent description
of known elementary particles and interactions.
The CERN Large Hadron Collider (LHC) has discovered
the Higgs boson to complete the SM field contents
\cite{Higgs1,Higgs2}.
Still the majority of matter in our Universe is, however, 
dark matter (DM) beyond the reach of our knowledge.
Thus the existence of hidden sectors is an exciting possibility
as an explanation of many problems beyond the SM
including DM. 

If DM is a fermion of the SM gauge singlet,
a mediator field would connect DM to the SM sector
with renormalizable couplings.
One of the minimal choice for the mediator field 
is a real singlet scalar
which is coupled to singlet fermionic dark matter (SFDM)
with the Yukawa type interaction
and to the SM through the quadratic term of the Higgs field,
the only massive coupling in the SM lagrangian.
Various aspects of such kind of minimal models, 
so called Higgs portal, 
has been studied in extensive literatures 
\cite{HPortal1,SFDM1,HPortal2,SFDM2,
HPortal3,HPortal4,HPortal5,HPortal6,HPortal6-1,
SFDM3,HPortal7,HPortal8,HPortal9,SFDM4,SFDM5,HPortal10}.
If there is a gauge symmetry in the hidden sector,
a vector field could be the mediator 
between the DM field in the hidden sector and the SM fields.
When the hidden gauge symmetry is U(1),
the corresponding gauge field can be coupled to the SM fields 
through the kinetic mixing with the field strength
of the SM U(1) gauge interaction.
Then the vector field has vectorlike couplings 
to the SM sector and is usually called a dark photon.
The hidden U(1) gauge symmetry is spontaneously broken 
in the hidden sector to yield the dark photon mass.

In this work, we consider an alternative way 
to connect the hidden sector including fermionic DM 
without the kinetic mixing to the SM.
We introduce an additional scalar field
which is the SM doublet and has the U(1)$_X$ charge
to connect the hidden U(1)$_X$ gauge field
to the SM fields.
The new scalar doublet does not couple to the SM fermions
due to the U(1)$_X$ charge, 
but couples to the SM Higgs doublet in the scalar potential
as well as the SU(2)$_L$ gauge fields.
Thus the Higgs sector is same 
as that of the two Higgs doublet model (2HDM) of type I.
It is pointed out in Ref. \cite{ko}
that an additional U(1) gauge symmetry
can explain the type I 2HDM flavour structure
instead of the discrete symmetry.
The U(1)$_X$ gauge boson gets the mass 
via the electroweak symmetry breaking (EWSB) in this model
and is mixed with the $Z$ boson. 
Since the new gauge boson, $Z'$, is mixed with only the $Z$ boson,
its couplings to the SM fermions are same as
the $Z$ boson couplings 
except for involving a suppression factor.
Thus we call it a dark $Z$ boson.
The dark $Z$ boson mass should be of the EW scale or less,
and actually expected to be much light. 
We anticipate that the couplings of the dark $Z$ 
to the SM should be very small due to constraints 
from lots of low energy neutral current (NC) experiments.
We consider the $\rho$-parameter, 
the atomic parity violation of Cs atom,
and the rare decays of $K$ and $B$ mesons
as experimental constraints in this work.
Note that the new gauge coupling need not be extremely small 
to suppress the dark $Z$ couplings to the SM sector,
if the Higgs doublet mixing $\sim 1/\tan \beta$ 
could be small enough.

SFDM carries the U(1)$_X$ charge 
and is connected to the SM through the dark $Z$ after the EWSB.
Since we have no restrictions on the U(1)$_X$ charge of SFDM,
the interaction strength of SFDM $\sim g_X X_\psi$
is a new free parameter to fit the observed relic density
and the DM-nucleon cross sections
under the bounds from direct detection experiments of DM.
We show that our SFDM mediated by the dark $Z$ 
can be a good DM candidate
satisfying the stringent experimental constraints on
the dark $Z$, DM and Higgs phenomenology.

This paper is organized as follows.
We describe the model in section 2.
Presented are the experimental constraints on the dark $Z$ boson 
from the $\rho$ parameter, 
the atomic parity violation of C$_s$ atom,
and decays of $K$ and $B$ mesons in section 3.
The dark matter phenomenology is studied in section 4
and the Higgs sector phenomenology in section 5.
We discuss the predictions for the future experiments
and conclude in section 6. 

\section{The Model}

We consider the QED-like hidden sector which consists of 
a SM gauge singlet Dirac fermion and the U(1)$_X$ gauge field.
No fields in the SM lagrangian carry the U(1)$_X$ gauge charge
and no kinetic mixing with the SM U(1)$_Y$ gauge field is assumed.
We introduce an additional scalar field as a mediator 
between the hidden sector and the visible sector,
which is the SM SU(2) doublet and carries the U(1)$_X$ charge.
The  charge assignment of two Higgs doublets $H_1$ and $H_2$, 
and the hidden fermion $\psi$ based on the gauge group 
${\rm SU}(3)_c \times {\rm SU}(2)_L \times 
{\rm U}(1)_Y \times {\rm U}(1)_X$
is given by
\be
H_1 (1, 2, \frac{1}{2}, \frac{1}{2}),~~~
H_2 (1, 2, \frac{1}{2}, 0),~~~
\psi (1, 1, 0, X),
\ee
where the U(1)$_X$ charge of $H_1$ is fixed to be 1/2
for convenience and that of $\psi$ is a free parameter.

Since the additional scalar doublet $H_1$ does not 
couple to the SM fermions due to the U(1)$_X$ charge,
the visible sector lagrangian of our model looks like
the 2HDM of type I 
except for the extra U(1)$_X$ gauge interaction for $H_1$.
We write the Higgs sector lagrangian as
\be
{\cal L}_H = (D^\mu H_1)^\dagger D_\mu H_1 
	   + (D^\mu H_2)^\dagger D_\mu H_2 - V(H_1, H_2) 
	   + {\cal L}_{\rm Y}(H_2), 
\ee
where $V(H_1, H_2)$ is the Higgs potential and 
${\cal L}_{\rm Y}$ the Yukawa interactions of the SM fermions.
The covariant derivative is defined by
\be
D^\mu = \partial^\mu + i g W^{\mu a} T^a 
                 + i g' B^\mu Y + i g_X A_X^\mu X,
\ee
where $X$ is the hidden U(1)$_X$ charge operator and 
the $A_X^\mu$ corresponding gauge field.
The Higgs potential is given by
\be
V(H_1,H_2) &=& \mu_1^2 H_1^\dagger H_1 + \mu_2^2 H_2^\dagger H_2
\nonumber \\
	   && + \lambda_1 (H_1^\dagger H_1)^2
	      + \lambda_2 (H_2^\dagger H_2)^2
              + \lambda_3 (H_1^\dagger H_1)(H_2^\dagger H_2)
              + \lambda_4 (H_1^\dagger H_2)(H_2^\dagger H_1).
\ee
Note that the $H_1^\dagger H_2$ quadratic term 
and the quartic term with $\lambda_5$ coupling are forbidden 
by the U(1)$_X$ gauge symmetry.

After the EWSB,
the vacuum expectation values (VEVs) of two Higgs doublets arise,
$\langle H_i \rangle = (0,v_i/\sqrt{2})^T$ with $i=1,2$,
and the gauge bosons get masses as
\be
{\cal L}_M = \frac{1}{4} g^2 v^2 W^+ W^-
             + \frac{1}{8} 
	       \left( \begin{array}{c} 
	              A_X \\[1pt] W^3 \\[1pt] B \end{array} \right)^T
	       \left( \begin{array}{ccc} 
                      g_X^2 v_1^2 &\ -g g_X v_1^2 &\ g' g_X v_1^2 \\[1pt]
		       -g g_X v_1^2 &\ g^2 v^2 &\ - g g' v^2 \\[1pt]
		       g' g_X v_1^2 &\ -g g' v^2 &\ {g'}^2 v^2 \\[1pt]
	              \end{array}
               \right)        
	       \left( \begin{array}{c} 
	              A_X \\[1pt] W^3 \\[1pt] B \end{array} \right),
\ee
where 
$v^2=v_1^2+v_2^2$. 
Diagonalizing the mass matrix 
with the Weinberg angle $\theta_W$ between $W^3$ and $B$,
we get the massless mode, the photon,
and diagonalization with the additional mixing angle $\theta_X$
between $A_X$ and the ordinary $Z$ mode
follows to get the physical masses such as,
\be
\left( \begin{array}{c}
        A_X \\[1pt] W^3 \\[1pt] B \end{array} \right)
= \left( \begin{array}{ccc}
                1 &\  0   &\ 0 \\[1pt]
                0 &\  c_W &\ s_W \\[1pt]
                0 &\ -s_W &\ c_W \\[1pt]
               \end{array}
        \right)
\left( \begin{array}{ccc}
                c_X &\  s_X &\ 0 \\[1pt]
               -s_X &\  c_X &\ 0 \\[1pt]
                0 &\ 0 &\ 1 \\[1pt]
               \end{array}
        \right)
        \left( \begin{array}{c}
              Z' \\[1pt] Z \\[1pt] A \end{array} \right)
= \left( \begin{array}{c}
              c_X Z' + s_X Z \\[1pt]  
             -s_X c_W Z' + c_X c_W Z + s_W A \\[1pt] 
	      s_X s_W Z' - c_X s_W Z + c_W A \end{array} \right).
\ee
where $s_W=\sin \theta_W=g'/\sqrt{g^2+{g'}^2}$,
$s_X=\sin \theta_X$ and 
\beq
\tan 2 \theta_X 
                = \frac{-2 g_X \sqrt{g^2 + {g'}^2} v_1^2}
                       {(g^2+{g'}^2)v^2 - g_X^2 v_1^2}
                = \frac{-2 g_X \sqrt{g^2 + {g'}^2} \cos^2 \beta }
                       {(g^2+{g'}^2) - g_X^2 \cos^2 \beta},
\eeq
with $\tan \beta =v_2/v_1$.
Since the kinetic mixing is ignored in this work,
the $Z-Z'$ mixing $\theta_X$ is originated from the mixing
in the mass matrix of Eq. (5).
Then the neutral gauge boson masses are
\be
m_{Z,Z'}^2 = \frac{1}{8} \left( g_X^2 v_1^2 +(g^2+{g'}^2) v^2
	       \pm \sqrt{(g_X^2 v_1^2 - (g^2+{g'}^2) v^2)^2
			 + 4 g_X^2 (g^2+{g'}^2) v_1^4} \right).
\ee
Note that only two mixing angles are required to diagonalize
the neutral gauge boson mass matrix in this model.  

We write the NC interactions in terms of the physical states
of the gauge bosons:
\be
{\cal L}_{NC} &\sim& 
    - e A^\mu \bar{f} Q \gamma_\mu f
    - c_X Z^\mu \left( g_L \bar{f}_L \gamma_\mu f_L
	  + g_R \bar{f}_R \gamma_\mu f_R \right)
\nonumber \\
   && + s_X {Z'}^\mu \left( g_L \bar{f}_L \gamma_\mu f_L
	  + g_R \bar{f}_R \gamma_\mu f_R \right),
\ee
where the electric charge is defined by $Q=T_3 + Y$ and
\be
e = \frac{gg'}{\sqrt{g^2+{g'}^2}},~~~~
g_L = -\frac{1}{2} \frac{g^2 - {g'}^2}{\sqrt{g^2+{g'}^2}},~~~~
g_R = \frac{{g'}^2}{\sqrt{g^2+{g'}^2}}.
\ee
Note that $g_L$ and $g_R$ are common with $Z'$ and $Z$
but the $Z'$ couplings involve 
the suppression factor, $-\sin \theta_X$.
This is the reason why we call $Z'$ the dark $Z$.

The structure of the Higgs sector is almost same as 
that of the type I 2HDM.
The only difference is that the pseudoscalar Higgs boson
does not exist in this model 
due to being the longitudinal mode of the dark $Z$.
Thus there are only  three additional Higgs bosons in this model,
a neutral CP-even Higgs boson and a pair of charged Higgs bosons.

The physical CP-even neutral Higgs bosons 
$h_1$, $h_2$ are defined by
\be
        \left( \begin{array}{c}
        \rho_1  \\[1pt]
        \rho_2  \\[1pt] \end{array}
        \right)
      = \left( \begin{array}{cc}
        \cos \alpha &\  \sin \alpha \\[1pt]
       -\sin \alpha &\  \cos \alpha \\[1pt] \end{array}
        \right)
        \left( \begin{array}{c}
        h_1  \\[1pt]
        h_2  \\[1pt] \end{array}
        \right)
      = \left( \begin{array}{c}
        h_1 \cos \alpha + h_2 \sin \alpha  \\[1pt]
       -h_1 \sin \alpha + h_2 \cos \alpha  \\[1pt] \end{array}
        \right),
\ee
where $\rho_i$ are the neutral components of
the doublets, $H_i=(H_i^+, (\rho_i+i \eta_i)/\sqrt{2})^T$,
and the mixing angle $\alpha$ is defined by
\be
\tan 2\alpha = \frac{(\lambda_3 + \lambda_4) \tan \beta}
                    {\lambda_1 - \lambda_2 \tan^2 \beta}.
\ee
The masses are obtained by
\be
M_{1,2}^2 = \lambda_1 v_1^2 + \lambda_2 v_2^2 \mp
           \sqrt{ (\lambda_1 v_1^2 - \lambda_2 v_2^2)^2
                  +(\lambda_3 + \lambda_4)^2 v_1^2 v_2^2}.
\ee
The heavier mode $h_2$ is the SM Higgs
and $h_1$ is the extra neutral Higgs boson
with relevant values of parameters
as will be shown later.

The charged Higgs boson masses are diagonalized
to get the physical mode $H^\pm$ by,
\be
        \left( \begin{array}{c}
        H_1^\pm  \\[1pt]
        H_2^\pm  \\[1pt] \end{array}
        \right)
      = \left( \begin{array}{cc}
        \cos \beta &\  \sin \beta \\[1pt]
       -\sin \beta &\  \cos \beta \\[1pt] \end{array}
        \right)
        \left( \begin{array}{c}
        G^\pm  \\[1pt]
        H^\pm  \\[1pt] \end{array}
        \right)
      = \left( \begin{array}{c}
        G^\pm \cos \beta + H^\pm \sin \beta  \\[1pt]
       -G^\pm \sin \beta + H^\pm \cos \beta  \\[1pt] \end{array}
        \right),
\ee
where the mixing angle is $\beta$ in this case.
One of the diagonalized masses is given by
\be
m_\pm^2 = -\frac{1}{2} \lambda_4 (v_1^2+v_2^2)
        = -\frac{1}{2} \lambda_4 v^2,
\ee
for $H^\pm$ and the other is 0 for $G^\pm$. 
The massless mode $G^\pm$ is the Goldstone mode
eaten up to be the longitudinal mode of the $W^\pm$ boson.
We write the Yukawa interactions for the charged Higgs boson 
with the short-hand notation
\be
{\cal L}_Y &=& - g_{ij}^d \bar{Q}^i_L H_2 d^j_R
              - g_{ij}^u \bar{Q}^i_L \tilde{H}_2 u^j_R
              - g_{ij}^l \bar{L}^i_L H_2 l^j_R   + H.C. ,
\nonumber \\
&=& - \frac{\sqrt{2} \cot \beta}{v} H^+ 
    \left( m_d   \bar{u}_L V_{\rm CKM} d_R 
          - m_u  \bar{u}_L V_{\rm CKM} d_R 
          - m_l  \bar{\nu}_L l^-_R 
    \right)
    + H.C.~,
\ee
where $V_{\rm CKM}$ are the corresponding quark mixings.

\section{Dark $Z$ phenomenology}

The NC interactions with the dark $Z$ boson
are constrained by various experiments.
Apart from the new Higgs masses and mixings,
the independent model parameters are $(g_X, \tan \beta)$
in our model lagrangian.
Instead in this analysis, we present the results
in terms of the observables
$(m_{Z'}, -s_X)$.

\subsection{The $\rho$ parameter}

We consider the precision test on the electroweak sector
using the $\rho$ parameter.
The $\rho$ parameter is defined 
by the ratio of $W$ and $Z$ boson masses,
$\rho \equiv m_W^2/m_Z^2 c_W^2$,
and should be 1 at tree level in the SM.
In this model, we have $m_W=gv/2$ as in the SM at tree level.
But the $Z$ boson mass is shifted such that
\be
m_Z^2 &=& \frac{m_W^2}{c_W^2 c_X^2} - m_{Z'}^2 \frac{s_X^2}{c_X^2},
\ee
and then the inverse of the $\rho$ parameter is
\be
\frac{1}{\rho} = \frac{m_Z^2 c_W^2}{m_W^2}
      = \frac{1}{c_X^2}
         - \frac{m_{Z'}^2 c_W^2}{m_W^2} \frac{s_X^2}{c_X^2}
      \approx 1 + s_X^2 \left(1 - \frac{m_{Z'}^2 c_W^2}{m_W^2} \right),
\ee
in the leading order of $s_X^2$.
The deviation $\Delta \rho$ from the unity is defined by
\be
\rho \equiv \frac{1}{1-\Delta \rho},
\ee
then the leading contribution to $\Delta \rho$ in this model 
is given by
\be
\Delta \rho_X 
     = -s_X^2 \left(1 - \frac{m_{Z'}^2 c_W^2}{m_W^2} \right).
\ee

\begin{figure}[t]
\centering
\includegraphics[width=12cm]{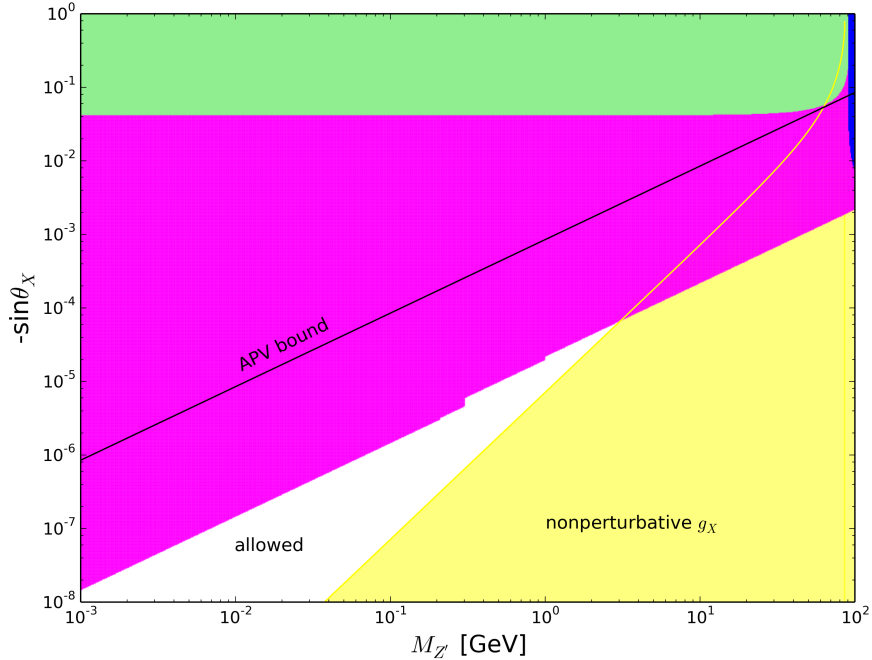}
\caption{Excluded regions in $(M_{Z'}, -s_X)$ plane.
The green (light grey) region is excluded by too small $\Delta \rho$
and the blue (dark grey) region excluded by too large $\Delta \rho$.
The pink (grey) region shows 
exclusions by rare $K$ and $B$ decays.
The solid line denotes the atomic parity violation bound
and the region above the line is excluded. 
The yellow region is disfavoured by the nonperturbative U(1)$_X$ gauge coupling, $g_X>4 \pi$. 
The allowed points of $(M_{Z'}, -s_X)$ fall on the white triangular region. 
}
\end{figure}

The correction $\Delta \rho$ is related to
$T$ parameter as \cite{peskin}
\beq
\Delta \rho = \alpha(m_Z)~ T
\eeq
of which values are 
\be
T = 0.07 \pm 0.12,
\ee
and ${\alpha^{(5)}}^{-1}(m_Z) = 127.955 \pm 0.010$
obtained in Ref. \cite{PDG}.
Then we have bounds for $\Delta \rho$ as
\be
-0.00039 < \Delta \rho < 0.001485.
\ee
Applying this bound to $\Delta \rho_X$, 
we show the excluded regions 
in $(m_{Z'}, -s_X)$ plane in Fig. 1.
The green (light grey) region denotes too small $\Delta \rho$
and the blue (dark grey) region  too large $\Delta \rho$.
Note that the yellow region denotes
the breakdown of the perturbativity, $g_X>4 \pi$.

\subsection{The atomic parity violation}

The parity violation of the atomic spectra is observed
due to the $Z$ boson exchanges.
The precise measurement of the atomic parity violation (APV) 
provides a strong constraint on the exotic NC interactions.
We derive the effective lagrangian 
for the corresponding process as
\be
-{\cal L} = -\frac{G_F}{\sqrt{2}} \left(
      g_{AV}^u ({\bar e} \gamma_\mu \gamma^5 e)({\bar u} \gamma^\mu u)
     + g_{AV}^d ({\bar e} \gamma_\mu \gamma^5 e)({\bar d} \gamma^\mu d)
	      \right),
\ee
at the quark level.

The APV is described by the weak charge of the nuclei defined by
\beq
Q_W \equiv -2 \left[ Z g_{AV}^p + N g_{AV}^n \right],
\eeq
where $Z$ ($N$) is the number of protons (neutrons) in the atom
and the nucleon couplings are defined by
$g_{AV}^p \equiv 2 g_{AV}^u+g_{AV}^d$ 
and $g_{AV}^n \equiv g_{AV}^u+2 g_{AV}^d$. 
In the SM,
$g_{AV}^p \approx -1/2 + 2 s_W^2$ and $g_{AV}^n \approx 1/2$ 
lead to $Q_W^{SM} \approx -N + Z (1-4 s_W^2)$ at tree level,
which is shifted by the dark $Z$ contribution as
\beq
Q_W = Q_W^{SM} \left( 1+\frac{m_Z^2}{m_{Z'}^2} s_X^2 \right),
\eeq
in the leading order of $s_X$.
The SM prediction of the Cs atom is
\cite{APVSM1,APVSM2}
\beq
Q_W^{SM} =  -73.16\pm0.05,
\eeq
and the present experimental value is 
\cite{APV}
\beq
Q_W^{exp} = -73.16\pm0.35,
\eeq
which yields the bound
\beq
\frac{m_Z^2}{m_{Z'}^2} s_X^2 \le 0.006,
\eeq
at 90 \% CL \cite{hslee}.
This constraint is shown as the solid line 
of the $(m_{Z'},-s_X)$ plane in Fig. 1.
The region above the line is excluded.

\subsection{Rare meson decays}

The flavour physics have been a good laboratory
of new physics.
Davoudiasl et al. \cite{hslee} suggest that
the flavour-changing neutral current (FCNC) decays 
of $K$ and $B$ mesons provide 
strong constraints on the dark $Z$ model.
Here, we follow their analysis to constrain our model.

The FCNC interactions of the dark $Z$ boson
$s \to d Z'$ and $b \to s Z'$ derive
$K \to \pi Z'$ and $B \to K (K^*) Z'$ decays,
\be
{\rm Br}(K^+ \to \pi^+ Z') &\approx& 
         4 \times 10^{-4} \left( \frac{m_Z}{m_{Z'}} \right)^2 s_X^2,
\nonumber \\
{\rm Br}(B \to K Z') &\approx& 0.1 
         \left( \frac{m_Z}{m_{Z'}} \right)^2 s_X^2,
\ee
and sequential decays of $Z'$ into lepton pairs lead to
rare decays $K \to \pi l \bar{l}$ and $B \to K l \bar{l}$.
The experimental measurements for $K$ mesons
\be
{\rm Br}(K^+ \to \pi^+ e^+ e^-) &=& (3.00\pm0.09) \times 10^{-7},
\nonumber \\
{\rm Br}(K^+ \to \pi^+ \mu^+ \mu^-) &=& (9.4\pm0.6) \times 10^{-8},
\nonumber \\
{\rm Br}(K^+ \to \pi^+ \nu^+ \nu^-) &=& (1.7\pm1.1) \times 10^{-10},
\ee
and for $B$ mesons
\be
{\rm Br}(B \to K l^+ l^-) &=& (4.51\pm0.23) \times 10^{-7},
\nonumber \\
{\rm Br}(B^+ \to K^+ \nu \nu) &<& 1.6 \times 10^{-5},
\ee
are obtained \cite{PDG}.
Then the strongest constraints are derived
\cite{hslee}
\be
\left| \frac{m_Z}{m_{Z'}} s_X \right| 
&\le& \frac{0.001}{\sqrt{{\rm Br}(Z' \to l^+ l^-)}},
\nonumber \\
\left| \frac{m_Z}{m_{Z'}} s_X \right| 
&\le& \frac{0.001}{\sqrt{{\rm Br}(Z' \to {\rm missing})}}.
\ee
Although being not manifest in the analysis,
the DM mass affects these constraints.
If the DM mass is less than the half of the $Z'$ mass,
the DM pair production channel opens and even dominates the decay rates,
${\rm Br}(Z' \to {\rm missing}) \sim$ 100$\%$ 
due to the $s_X^2$ suppression of the $Z'$ decays into the SM final states.  
Then the ${\rm Br}(K^+ \to \pi^+ l^+ l^-)$ and
${\rm Br}(B \to K l^+ l^-)$ constraints do not work.
The pink (grey) region in Fig. 1 denotes the excluded points
by the constraints given in Eq. (33).

The final result is depicted in Fig. 1,
where constraints from $\Delta \rho$, APV, and rare meson decays
are presented altogether.
We find that the rare meson decays provide
the strongest constraints on $m_{Z'}$ and $|s_X|$.
We also see that the dark $Z$ is rather light, $m_{Z'} \le 2$ GeV,
and the coupling should be very small due to the small mixing angle 
$|\sin \theta_X| \le 4 \times 10^{-5}$ as expected.

We note that the sign of $\theta_X$ is not determined 
by the phenomenological study of this section.
At this stage, we just know that only very small $|\theta_X|$
are allowed by the experiments.
Then we can see that $g_X$ or $ \cos \beta$ should be small in the Eq. (7)
for $|\theta_X|$ to be very small, $< 10^{-5}$
and then $\theta_X$ should be negative. 
As we see that the allowed region of Fig. 1 is 
near the nonperturbativity region of $g_X$,
actually only the small $ \cos \beta$ (means very large $\tan \beta$)
is allowed by our analysis.

\section{Dark Matter Phenomenology}

\begin{figure}[t]
\centering
\includegraphics[width=12cm]{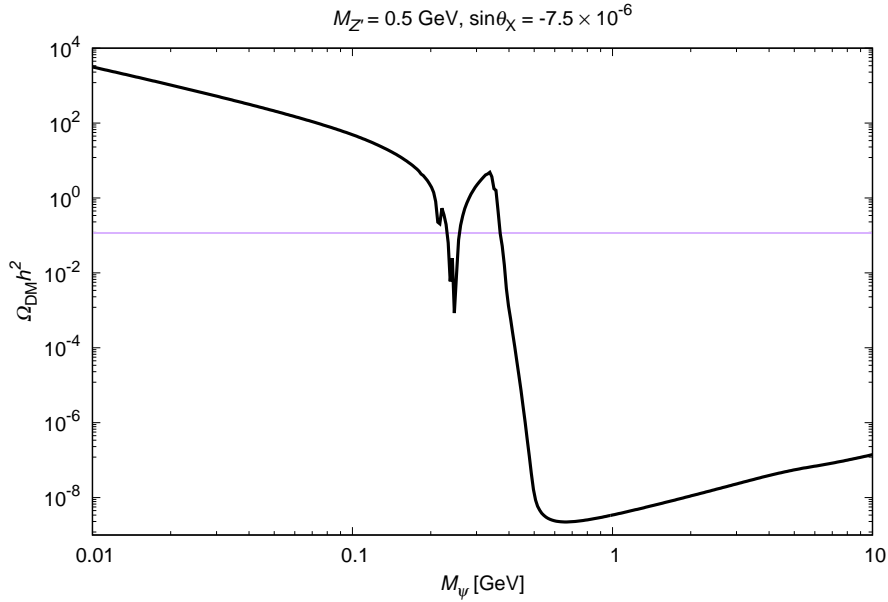}
\caption{The relic density with respect to the DM mass 
for the benchmarking point 
$(m_{Z'},\sin \theta_X) = (0.5~{\rm GeV}, -7.5 \times 10^{-6})$.
}
\end{figure}

Our hidden sector consists of a Dirac fermion 
with a U(1)$_X$ gauge symmetry.
The hidden sector lagrangian is QED-like
\be
{\cal L}_{\rm hs} = -\frac{1}{4} F_X^{\mu \nu} F_{X \mu \nu} 
                    + \bar{\psi} i \gamma^\mu D_\mu \psi
                          - M_\psi \bar{\psi} \psi,
\ee
where
\be
D^\mu = \partial^\mu + i g_X A^\mu_X X,
\ee
and $X$ is the U(1)$_X$ charge operator for $\psi$.
We show that the singlet fermion $\psi$ can be a DM candidate.
Using Eq. (6), 
we find that $\psi$ has vectorial interactions with $Z$ and $Z'$ bosons, 
\be
{\cal L}_{\rm DM}^{int} = i g_X ~X~ \bar{\psi} \gamma^\mu \psi
			  \left( c_X {Z'}_\mu + s_X Z_\mu \right).
\ee
We have two additional parameters, $m_\psi$ 
and the U(1)$_X$ charge
for the DM phenomenology.
Since the couplings between the DM sector and the SM sector involves
a suppression factor $|s_X|$,
the collider phenomenology including DM is affected very little
with the $s_X^2$ suppression.

The SFDM contribution to the relic abundance density $\Omega$ 
is obtained from global fits of 
various cosmological observations.
We can read the present value of $\Omega$ 
of the cold nonbaryonic DM as
\beq
\Omega_{\rm CDM} h^2 = 0.1186 \pm 0.0020,
\eeq
from measurements of the anisotropy 
of the cosmic microwave background (CMB)
and of the spatial distribution of galaxies
\cite{PDG}.
Such precise value provides a stringent constraint
on the model parameters.
We calculate $\Omega$ and the DM-nucleon cross section 
using the \texttt{micrOMEGAs} \cite{micromegas}
with the allowed values of parameters $(m_{Z'}, \sin \theta_X)$
given in the previous section.
Figure 2 shows the relic density with respect to the DM mass 
for the benchmarking point $m_{Z'} = 0.5$ GeV and
$\sin \theta_X = -7.5 \times 10^{-6}$.
The acceptable DM annihilations for the relic abundance arise
at the resonant region where $M_\psi \sim m_{Z'}/2$
through the $s$-channel 
$\psi \bar{\psi} \to Z' \to $ SM particles
and at the nonresonant region 
through the $t$-channel $\psi \bar{\psi} \to Z' Z'$ and 
the Higgsstrahlung $\psi \bar{\psi} \to Z' h_1 $ processes.

\begin{figure}[t]
\centering
\includegraphics[width=12cm]{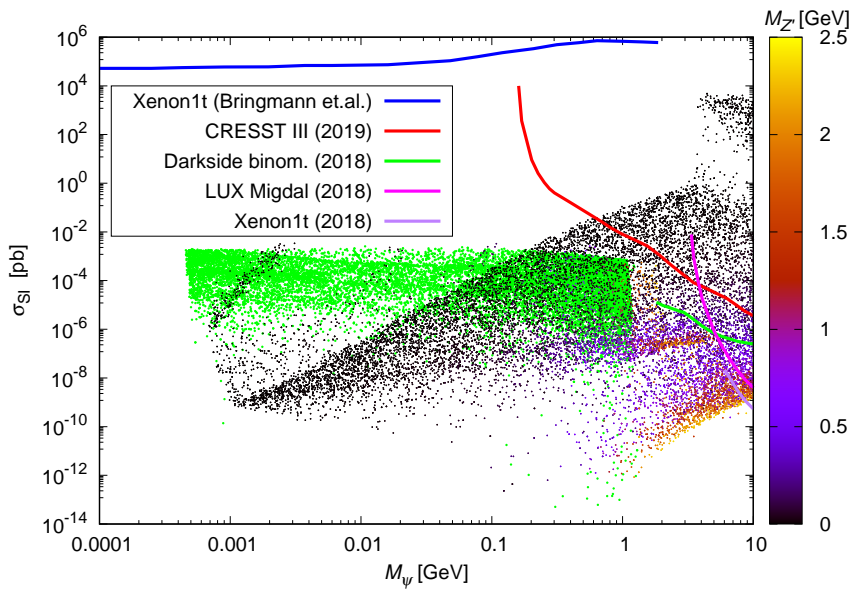}
\caption{
The DM-nucleon cross sections for the parameter sets
satisfying the relic abundance. 
Green points denote the resonant region for DM annihilation
and points of other colors the nonresonant annihilations
depending on the $Z'$ and $h_1$ masses.
}
\end{figure}

The direct detection cross sections for SFDM are calculated 
for the parameters satisfying the relic abundance of Eq. (37) 
and shown with respect to the DM mass in Fig. 3.
We can see two groups of allowed points in the plot.
The green points denote the resonant annihilations,
hence $M_\psi \sim m_{Z'}/2$ and are distributed
in the region of $M_\psi < 1.2$ GeV.
Points of other colors for the nonresonant annihilations
are distributed in the whole region of $M_\psi$,
but are excluded by the present experiments
when $M_\psi > 10$ GeV.
The experimental bounds from Xenon1t \cite{Xenon1t}, 
CRESST III \cite{CRESST}, Darkside \cite{Darkside}, and
LUX \cite{LUX} are shown together.

We have to mention that DM also interacts with the SM
through the ordinary $Z$ boson.
For the contributions to the relic density
and the DM-nucleon cross sections,
the suppression factor for coupling strengths
are same order for both $Z$ and $Z'$ mediation.
Thus the dark $Z$ mediation is dominant 
when the DM mass is around the dark $Z$ mass,
and the ordinary $Z$ mediation dominant
when the DM mass around $m_Z$.
However, 
the $Z$ mediated contributions are excluded by
the stringent experimental bound 
when the DM mass is a few tens GeV.

Production of energetic particles 
due to self-annihilation of DM 
in high DM density regions like galactic center
has been studied by several telescopes
as the indirect signal of DM.
The present observations provide constraints
on the velocity-weighted annihilation cross sections
for various channels, but among them, 
the $\tau^+ \tau^-$ or $b\bar{b}$ channels
are strongly constrained by energetic photon spectrum data.
We take a benchmarking point with values
$M_\psi \approx 2.1$ GeV, $m_{Z'} \approx 0.15$ GeV,
$-\sin \theta_X \approx 1.1 \times 10^{-6}$,
which satisfies the observed relic abundance
and gives the DM-nucleon cross section of order $10^{-6}$ pb. 
Since this DM fermion is rather light,
the $\tau^+ \tau^-$ channel is more relevant.
The annihilation rate for the $\tau^+ \tau^-$ channel is
$\langle \sigma v \rangle \approx 1.8 \times 10^{-37}$ cm$^{3}$s$^{-1}$
much below the observed bounds $\sim 10^{-27}$ cm$^{3}$s$^{-1}$
\cite{FermiLAT,HESS}.
We survey the DM mass, $2 < M_\psi <3$ in GeV,
and find that the annihilation rate is generically small, 
$\langle \sigma v \rangle < 10^{-34}$ cm$^{3}$s$^{-1}$
due to the small mixing $|s_X|$.
Thus our model is safe for the present bounds
from the indirect search of DM.

\section{Higgs Phenomenology}

An additional Higgs doublet that is chraged 
under the U(1)$_X$ symmetry is introduced 
and extra scalar particles exist in our model.
This set-up is similar to the Type-I 2HDM 
with U(1)$_H$ gauge symmetry in Ref.~\cite{pkotype1}.
We note that in both models the quartic term 
with the $\lambda_5$ coupling is forbidden due to
additional gauge symmetry. 
However, by introducing another singlet field $\Phi$,
the latter is allowed to have the $H_1^\dagger H_2$ quadratic term 
which can make extra scalars relatively heavy, whereas, in our model, 
the masses of extra scalar bosons are bounded by the perturbativity condition, 
for example, 
$m_{\pm}^2 = - \frac{1}{2} \lambda_4 v^2 \lesssim (616~\rm{GeV})^2$. 
In our model the CP-odd scalar mode is eaten up
by the dark $Z$ boson and there exists no CP-odd scalar.
This is the noticeable difference from the particle contents in the ordinary 2HDM 
as well as in the 2HDM with U(1)$_H$ gauge symmetry~\cite{pkotype1}.
As a result the new particles are an neutral Higgs boson
and a pair of charged Higgs bosons.
Most of the phenomenology of the Higgs sector is governed
by quartic couplings of the Higgs potential.
Hence we just discuss two issues on the Higgs phenomenology,
the charged Higgs search and the Higgs invisible decays here.

To begin with we investigate the scalar masses.
The masses are calculated
with the perturbativity conditions
on the quartic couplings $|\lambda_i|<4 \pi$
and the vacuum stability conditions \cite{pkotype1},
\be
\lambda_1 > 0,~~
\lambda_2 > 0,~~
\lambda_3 > -2\sqrt{\lambda_1 \lambda_2},~~
\lambda_3 + \lambda_4 > -2\sqrt{\lambda_1 \lambda_2}.
\ee
We find that $h_1$ is very light in this model
since $v_1 \ll v_2$.
Hence $h_2$ should be the SM Higgs boson.
If we fix the mass of $h_2$ to be $125.18 \pm 0.16$ GeV,
the $h_1$ mass is less than 1.2 GeV.
The charged Higgs boson mass is determined
by $\lambda_4$ solely in this model
and has the upper bound $\sim 616$ GeV
due to the perturbativity bound of $\lambda_4$.
We note that
these features are very insensitive to the parameter set
allowed in the previous analysis.

For the analysis of $\Delta \rho$ in the previous section, 
we consider only the dark $Z$ contributions. 
By the way, the additional Higgs bosons also contribute
to the $\rho$ parameter such as \cite{hollik}
\be
\Delta \rho^{(1)}_{\rm NS} = \frac{\alpha}{16 \pi m_W^2 s_W^2}
     \left( m_\pm^2 - \frac{m_1^2 m_\pm^2}{m_1^2-m_\pm^2} 
              \log \frac{m_1^2}{m_\pm^2} 
        \right),
\ee
where $m_1$ is the mass of $h_1$
and $m_\pm$ the charged Higgs boson mass.
Since $h_1$ is very light compared with $H^\pm$,
$\Delta \rho^{(1)}_{\rm NS}$ crucially depends 
only on the charged Higgs mass.
If $m_\pm \ge 120$ GeV,
$\Delta \rho_{\rm NS}^{(1)}$ exceeds 0.001485 of
the experimental upper limit given in Eq. (23)
and no parameter set can satisfy the $\Delta \rho$.
On the other hand, $\Delta \rho_{\rm NS}^{(1)}$ 
is very sensitive to $m_\pm$
and it does not play a role of constraints 
if $m_\pm$ is just slightly smaller than 120 GeV, e.g. 119 GeV.
(The cyan region of Fig. 1 is overlapped by other constraints.)
Therefore we demand $80~{\rm GeV} < m_\pm < 120~ {\rm GeV}$ 
in this model. 
The model-independent lower bound is given in
\cite{PDG}.

The recent CMS data for $H^\pm \to \tau^\pm \nu$ and
$H^+ \to t \bar{b}$ channels at $\sqrt{s}=13$ TeV
with an integrated luminosity of 35.9 fb$^{-1}$
have been reported in Ref. \cite{CMS1,CMS2}.
Sanyal \cite{sanyal} provides the analysis of the CMS data
and shows the allowed values of $\tan \beta $
and the charged Higgs mass $m_\pm$ 
for several versions of the 2HDM including type I. 
Since our charged Higgs is rather light, $m_\pm<120$ GeV,
it is produced from the top decay and 
decays into $\tau \nu$ and $W^\pm h_1$.   
Figures 3 and 5 in Ref. \cite{sanyal} depict the exclusion regions
of $(m_\pm, \tan \beta)$ in the 2HDM of type I
from the upper limits of the CMS observations 
$\sigma_{H^\pm} {\rm Br}(H^\pm \to \tau^\pm \nu)$
and ${\rm Br}(B \to X_s \gamma$ constraints.
We find the conservative limit is $\tan \beta > 15$ 
for any values of $m_\pm$ from these plots. 
Meanwhile the parameter sets in the allowed region given in Fig. 1
correspond to very large $\tan \beta$, 
numerically $\tan \beta > 500$ for all $m_\pm$,
much larger than the CMS conservative limit.
Therefore the present LHC bound for $H^\pm$
is not relevant to our model.

Since the dark $Z$ boson is light in this model,
the Higgs boson decays into the dark $Z$ pair are possible
which contributes to the Higgs invisible decay modes.
However, the $h_2 Z' Z'$ coupling is suppressed by
$\sin^2 \theta_X$ or $g_X^2 \cos \beta \sin \alpha$.
Since $\sin \theta_X \sim g_X \cos^2 \beta$
and $\sin \alpha \sim \cos \beta$,
the decay rate $\Gamma(h \to Z' Z')$ is suppressed
by the factor $\sin^2 \theta_X$ or less
compared with $\Gamma(h \to Z Z)$.
Thus the $h \to Z' Z'$ contribution to the Higgs invisible decay
is much smaller than the current limit
Br$(h \to {\rm invisible}) < 0.22$ by the CMS
\cite{CMSinvisible}.

\section{Concluding Remarks}

We have constructed the SFDM model mediated by the dark $Z$ boson.
The hidden U(1) gauge boson does not couple to the SM sector directly
in this model, but interacts with the SM through the Higgs mixing
with an additional Higgs doublet involving the hidden U(1) charge. 
The Higgs mixing induces the $Z-Z'$ mixing,
and the mixing angle depends upon the Higgs mixing angle $\beta$
and the hidden gauge coupling $g_X$.
The dark $Z$ boson is severely constrained by the electroweak data
and thus the $Z-Z'$ mixing angle $\theta_X$ should be very small,
$|\sin \theta_X| < 4 \times 10^{-5}$.
The allowed parameter space by the experiments is corresponding to
the very large $\tan \beta$ region.
The mass of the dark $Z$ is approximately
the VEV of the additional Higgs doublet $v_1 = v \cos \beta $
and consequently it is rather light, less than 2 GeV.
Such a dark $Z$ boson could also affect the precision QED  tests. 
With values of  the allowed $(m_{Z'}, -\sin \theta_X)$ points in our analysis,
additional contributions of the dark $Z$ 
to the anomalous magnetic moments of the electron and the muon are negligible 
compared with the limit given in Ref. \cite{pospelov}.

In this model, our DM is a SM singlet fermion
and mediated by the dark $Z$ boson.
We find that it can satisfy the observed 
relic abundance from the CMB observation.
Since the dominant channels of the DM annihilation
in the early universe
are the $s-$channel at the dark $Z$ resonance region,
the $t-$channel at the $\psi \bar{\psi} \to Z' Z'$ opening region,
and the Higgsstrahlung into $Z' h_1$ region,
the DM mass is same order as the dark $Z$ mass, 
$\sim$ GeV and less.

The dark $Z$ boson might live long
if the DM mass is larger than the half of the dark $Z$ mass,
(Actually in that case, the DM mass is almost same as the dark $Z$ mass
to satisfy the relic density.)
since the coupling strength of the dark $Z$ to the SM matter is very small.
Then the proposed intensity frontier experiments,
e.g. SHiP \cite{SHiP}, FASER \cite{FASER}, 
MATHUSLA \cite{MATHUSLA} and etc.
will have the chance to probe the dark $Z$ boson directly
in the future.

\acknowledgments
This work is supported 
by Basic Science Research Program
through the National Research Foundation of Korea (NRF)
funded by the Ministry of Science, ICT, and Future Planning 
under the Grants No. NRF-2017R1E1A1A01074699 (DWJ, ShN),
and No. NRF-2017R1A2B4011946 (CY), No. NRF-2020R1A2C3009918 (CY),
and also funded by the Ministry of Education
under the Grants No. NRF-2018R1D1A1B07047812 (DWJ),
No. NRF-2018R1D1A1B07050701 (YGK),
and No. NRF-2018R1D1A3B07050649 (KYL).

\def\PRDD #1 #2 #3 {Phys. Rev. D {\bf#1},\ #2 (#3)}
\def\PRD #1 #2 #3 #4 {Phys. Rev. D {\bf#1},\ No. #2, #3 (#4)}
\def\PRLL #1 #2 #3 {Phys. Rev. Lett. {\bf#1},\ #2 (#3)}
\def\PRL #1 #2 #3 #4 {Phys. Rev. Lett. {\bf#1},\ No. #2, #3 (#4)}
\def\PLB #1 #2 #3 {Phys. Lett. B {\bf#1},\ #2 (#3)}
\def\NPB #1 #2 #3 {Nucl. Phys. B {\bf #1},\ #2 (#3)}
\def\ZPC #1 #2 #3 {Z. Phys. C {\bf#1},\ #2 (#3)}
\def\EPJ #1 #2 #3 {Euro. Phys. J. C {\bf#1},\ #2 (#3)}
\def\JPG #1 #2 #3 {J. Phys. G: Nucl. Part. Phys. {\bf#1},\ #2 (#3)}
\def\JHEP #1 #2 #3 {JHEP {\bf#1},\ #2 (#3)}
\def\JCAP #1 #2 #3 {JCAP {\bf#1},\ #2 (#3)}
\def\IJMP #1 #2 #3 {Int. J. Mod. Phys. A {\bf#1},\ #2 (#3)}
\def\MPL #1 #2 #3 {Mod. Phys. Lett. A {\bf#1},\ #2 (#3)}
\def\PTP #1 #2 #3 {Prog. Theor. Phys. {\bf#1},\ #2 (#3)}
\def\PR #1 #2 #3 {Phys. Rep. {\bf#1},\ #2 (#3)}
\def\RMP #1 #2 #3 {Rev. Mod. Phys. {\bf#1},\ #2 (#3)}
\def\PRold #1 #2 #3 {Phys. Rev. {\bf#1},\ #2 (#3)}
\def\IBID #1 #2 #3 {{\it ibid.} {\bf#1},\ #2 (#3)}

\end{document}